\documentstyle[12pt]{article}

\begin{document}
\vspace{8cm}
\begin{center}\begin{Large}\begin{bf} ADJOINT  QCD$_2$ IN LARGE N
\footnote{Talk presented at the 1997 Orbis Scientiae Conference Miami
Beach Florida, January 25-28,1997; based on work done with F. Antonuccio}

\end{bf}\end{Large}\end{center}
\vspace{.75cm}\begin{center}
Stephen Pinsky\\[10pt]
     
     \vspace{0.1cm}
     \begin{it}
Department of Physics\\
The Ohio State University\\
174 West 18th Avenue\\
Columbus, Ohio  43210\\
       \end{it}
         \end{center}
\vspace{1cm} \baselineskip=35pt \begin{abstract} \noindent
 We consider a dimensional reduction of 3+1 dimensional $SU(N)$
Yang-Mills theory coupled to adjoint fermions to obtain a class of
$1+1$ dimensional gauge theories. We derive the quantized
light-cone Hamiltonian in the light-cone gauge $A_- = 0$ and large-$N$
limit, then solve for the masses, wavefunctions and of the color
singlet boson and fermion boundstates. We find that the theory has many
exact massless state that are similar to the t'Hooft pion.
\end{abstract}

{\bf INTRODUCTION}
\vspace{.1in}
\baselineskip=18pt

In this  work \cite{anp97}, we start by considering
${\mbox{QCD}}_{3+1}$  coupled to Dirac adjoint fermions. Here, the
virtual creation of fermion-antifermion pairs is not suppressed in the
large-$N$ limit -- in contrast to the case for fermions in the
fundamental representation
\cite{and96b} -- and so one may study the structure of boundstates
beyond the valence quark (or quenched) approximation{\cite{anp97}. We
also anticipate that the techniques employed here will have special
interest in the context of solving supersymmetric matrix theories.

The ${\mbox{QCD}}_{3+1}$  theory coupled to adjoint fermions is
reduced to a $1+1$ dimensional field theory by stipulating that all
fields are independent of the transverse coordinates
$x^{\perp}=(x^1,x^2)$. The resulting theory is
${\mbox{QCD}}_{1+1}$  coupled to two $1+1$ dimensional complex adjoint
spinor fields, and two real adjoint scalars.  A key strategy in
formulating this model field theories is to retain as many of the
essential degrees of freedom of higher dimensional QCD while still
being able to extract complete non-perturbative solutions. One finds
Yukawa interactions between the scalars and fermion fields. While this
approach is not equivalent to solving the full
$3+1$ theory and then going to the regime where $k_\perp$ is
relatively small, it may share many qualitative features of the higher
dimensional theory, since the longitudinal dynamics is treated
exactly. Studies of this type for pure glue and with fundamental
quarks have yielded a number of interesting results
\cite{and96b,and96a,buv95}.

The unique features of light-front quantization
\cite{dir49} make it a powerful tool for the
nonperturbative study of quantum field theories.  The main advantage
of this approach is the apparent simplicity of the vacuum state. 
Indeed, naive kinematic arguments suggest that the physical vacuum is
trivial on the light front. Since in this case all fields transform in
the adjoint representation of $SU(N)$, the gauge group of the theory
is actually $SU(N)/Z_N$, which has nontrivial topology and vacuum
structure. For the particular gauge group
 $SU(2)$ this has been discussed elsewhere \cite{pir96}. While this
vacuum structure may in fact be relevant for a discussion on
condensates, for the purposes of this calculation they will be ignored.

In the first section we formulate the $3+1$ dimensional $SU(N)$
Yang-Mills theory and then perform dimensional reduction to obtain a
$1+1$ dimensional matrix field theory. The light-cone Hamiltonian is
then derived for the light-cone gauge $A_- = 0$ following a discussion
of the physical degrees of freedom of the theory. Singularities from
Coulomb interactions are regularized in a natural way, and we outline
how particular ``ladder-relations'' take care of potentially troubling
singularities for vanishing longitudinal momenta $k^+ = 0$. In the final
section exact massless solutions of the boundstate integral
equations are discussed. 

\vskip.3in
{\bf DEFINITIONS}
\vskip.1in

We first consider $3+1$ dimensional $SU(N)$ Yang-Mills coupled to
a Dirac spinor field whose components transform in the adjoint
representation of $SU(N)$:
\begin{equation}
{\cal L} = \mbox{Tr} \left[ -\frac{1}{4} F_{\mu \nu} F^{\mu \nu}
 + \frac{{\rm i}}{2}
(\bar\Psi\gamma ^{\mu} \buildrel \leftrightarrow \over
D_{\mu}\Psi) - m\bar\Psi \Psi \right] \; ,
\label{3+1theory}
\end{equation}
where $D_{\mu} = \partial_{\mu} + {\rm i}g [ A_\mu,\ \ ] $ and $F_{\mu \nu}
= \partial_{\mu}A_{\nu} - \partial_{\nu} A_{\mu} + {\rm i}g [A_{\mu},
A_{\nu} ]$. We also write $A_{\mu} = A_{\mu}^a \tau ^a$ where $\tau^a$ is
normalized such
that $\mbox{Tr} (\tau^a \tau^b ) = \delta_{ab}$.
The projection operators\footnote{
We use the conventions $\gamma^{\pm} = (\gamma^0 \pm \gamma^3)/
\sqrt{2}$, and $x^{\pm}=(x^0 \pm x^3)/\sqrt{2}$.}
$\Lambda_L,\Lambda_R$ permit a decomposition
of the spinor field $\Psi = \Psi_L +\Psi_R$, where
\begin{equation}
\Lambda_L = {1 \over 2} \gamma^+ \gamma^- ,\quad \Lambda_R = {1 \over 2}
\gamma^- \gamma^+ \quad \mbox{and} \quad
\Psi_L = \Lambda_L \Psi, \quad \Psi_R = \Lambda_R \Psi .
\end{equation}
Inverting the equation of motion for $\Psi_L$, we find
\begin{equation}
\Psi_L ={1 \over 2{\rm i}D_-} \left[ {\rm i}
\gamma^i D_i +m \right ] \gamma^+ \Psi_R
\label{eqnmotion}
\end{equation}
where $i=1,2$ runs over transverse space. Therefore $\Psi_L $ is not an
independent degree of freedom.

Dimensional reduction of the $3+1$ dimensional Lagrangian
(\ref{3+1theory}) is performed by assuming (at the classical level)
that all fields are independent of the transverse coordinates
$x^{\perp}=(x^1,x^2)$: $\partial_{\perp} A_{\mu} =0$ and
$\partial_{\perp} \Psi = 0$.
In the resulting $1+1$ dimensional field theory, the transverse components
$A_{\perp} = (A_1,A_2)$ of the gluon field will be represented
by the $N \times N$ complex matrix fields $\phi_{\pm}$:
\begin{equation}
     \phi_{\pm} = \frac{A_1 \mp {\rm i} A_2}{\sqrt{2}}.
\end{equation}
Here,  $\phi_-$ is just the Hermitian conjugate of
$\phi_+$. When the theory is quantized, $\phi_{\pm}$ will correspond to
$\pm 1$ helicity bosons (respectively).

The components of the Dirac spinor $\Psi$ are the $N \times N$
{\em complex} matrices $u_{\pm}$ and $v_{\pm}$, which
are related to the left and right-moving spinor fields according
to
\begin{equation}
\Psi_R ={1 \over 2^{{1 \over 4}}}
\left ( \begin{array}{c} u_+\\ 0 \\ 0 \\u_- \end{array}
\right )
\quad
\Psi_L ={1 \over 2^{{1 \over 4}}}
\left ( \begin{array}{c} 0 \\ v_+ \\ v_- \\0 \end{array}
\right )
\end{equation}
Adopting the light-cone gauge $A_- = 0$ allows one to explicitly
rewrite the left-moving fermion fields $v_{\pm}$ in terms
of the right-moving fields $u_{\pm}$ and boson fields $\phi_{\pm}$,
by virtue of equation (\ref{eqnmotion}). We may therefore eliminate
$v_{\pm}$ dependence from the field theory. Moreover,
Gauss' Law
\begin{equation}
\partial_-^2 A_+ = g \left( {\rm i}[\phi_+, \partial_- \phi_-] +
                    {\rm i}[\phi_-, \partial_- \phi_+]
 + \{ u_+,u_+^{\dagger} \} + \{ u_-,u_-^{\dagger} \} \right)
\end{equation}
permits one to remove any explicit dependence on $A_+$,
and so the remaining {\em physical}
degrees of freedom of the field theory are represented by
the helicity $\pm \frac{1}{2}$ fermions $u_{\pm}$, and the
helicity $\pm 1$ bosons $\phi_{\pm}$. There are no ghosts
in the quantization scheme adopted here.
In the light-cone frame
the Poincar\'e generators $P^-$ and $P^+$ for the
reduced $1+1$ dimensional field theory are given by
\begin{equation}
P^+=\int ^\infty _{-\infty} dx^- \mbox{Tr}  \biggl[
2 \partial_- \phi_- \cdot \partial_- \phi_+
+ {{\rm i} \over 2} \sum_h \left ( u^\dagger_h \cdot
\partial_- u_h -\partial_-
u^\dagger_h \cdot u_h \right )
\biggr ]
\end{equation}
\begin{equation}
P^- = \int ^\infty _{-\infty} dx^- \mbox{Tr} \biggl[
 m_b^2 \phi_+ \phi_-
-{ g^2 \over 2}{J}^+ {1\over \partial_-^2 }{J}^+ +
{ t g^2 \over 2} \left [\phi_+ , \phi_- \right ]^2
+\sum_h{F}^\dagger_h {1 \over {\rm i}\partial_-} {F}^\dagger_h
\biggr ]
\label{hamiltonian}
\end{equation}
where the sum $\sum_h$ is over $h=\pm$ helicity labels,
and
\begin{eqnarray}
J^+ & = & {\rm i}[\phi_+, \partial_- \phi_-] +
                    {\rm i}[\phi_-, \partial_- \phi_+]
 + \{ u_+,u_+^{\dagger} \} + \{ u_-,u_-^{\dagger} \} \\
F_{\pm} & = & \mp s g\left[ \phi_{\pm} \; , u_{\mp} \right ] +
{ m \over \sqrt{2}} u_{\pm} \label{Fterm}
\end{eqnarray}
We have generalized
the couplings
by introducing the variables $t$ and $s$, which do not
spoil the $1+1$ dimensional gauge invariance of the reduced theory;
the variable $t$ will determine the strength of the quartic-like interactions,
and the variable $s$ will determine the
strength of the Yukawa interactions between the fermion and boson fields,
and appears explicitly in equation (\ref{Fterm}). The dimensional
reduction of the original $3+1$ dimensional theory yields the canonical
values $s=t=1$.

Renormalizability of the reduced theory also requires
the addition of a
bare coupling $m_b$, which leaves
the $1+1$ dimensional gauge invariance intact.
In all calculations, the renormalized
boson mass ${\tilde m}_b$ will be set to zero.

Canonical quantization of the field theory
is performed by decomposing the boson
and fermion fields into Fourier expansions
at fixed light-cone time $x^+ = 0$:
\begin{equation}
u_\pm  = {1 \over  \sqrt {2\pi}} \int_{-\infty}^\infty dk
\hspace{1mm} b_\pm(k)  e^{-{\rm i}k x^-} \quad \mbox{and} \quad
\phi_{\pm} = {1 \over  \sqrt {2\pi}} \int_{-\infty}^\infty { dk \over
\sqrt{2|k|}}\hspace{1mm}a_{\pm}(k) e^{-{\rm i}k x^-}
\end{equation}
where $b_{\pm} = b_{\pm}^a\tau^a$ etc.
We also define
\begin{equation}
\quad b_{\pm}(-k) =d_{\mp}^{\dagger}(k),
\quad a_\pm(-k) = a_{\mp}^\dagger (k),
 \label{note}
\end{equation}
where $d_{\pm}$
correspond to antifermions.
Note that
$(b^{\dagger}_{\pm})_{ij}$
should be distinguished
from $b_{\pm ij}^{\dagger}$, since in the former the quantum conjugate
operator $\dagger$ acts on (color) indices, while
it does not in the latter. The latter formalism
is sometimes customary in the study of matrix models.
The precise connection between the usual gauge theory and matrix theory
formalism may be stated as follows:
\[
 b_{\pm ji}^{\dagger} =
b_{\pm}^{a\dagger}\tau^{a*}_{ji}=b_{\pm}^{a\dagger}\tau^a_{ij} =
(b_{\pm}^{\dagger})_{ij}
\]
The commutation and anti-commutation relations (in matrix formalism)
for the boson and fermion fields take the following
form in the large-$N$ limit ($k,{\tilde k} >0$; $h,h'={\pm}$):
\begin{equation}
\left [ a_{h ij}(k),a_{h' kl}^{\dagger}({\tilde k}) \right] =
\{ b_{h ij}(k), b_{h' kl}^{\dagger}({\tilde k}) \}
= \{ d_{h ij}(k), d_{h' kl}^{\dagger}({\tilde k}) \}
= \delta _{h h'} \delta_{jl}\delta_{ik} \delta(k-{\tilde k}),
\label{rhccrs}
\end{equation}
where have used the relation
$\tau^a_{ij} \tau^a_{kl} =
\delta_{il} \delta_{jk} -{1 \over N} \delta_{ij} \delta_{kl}$.
All other (anti)commutators vanish.

The Fock space of physical states
is generated by the color singlet states, which have a natural
`closed-string' interpretation. They are formed by
a color trace  of the fermion, antifermion and boson
operators
acting on the vacuum state $|0\rangle$.
Multiple string states couple to the theory with
strength $1/N$,
and so may be ignored.

\vskip.3in
{\bf THE LIGHT CONE HAMILTONIAN}
\vskip.1in

For the special case ${\tilde m}_b=m=t=s=0$, the light-cone Hamiltonian
is simply given by the current-current term $J^+ \frac{1}{\partial_-^2} J^+$
in equation (\ref{hamiltonian}). In momentum space,
this Hamiltonian takes the form
\begin{eqnarray}
P^-_{J^+ \cdot J^+} &=& {g^2 \over 2 \pi}
\int^\infty_{-\infty} dk_1 dk_2 dk_3 dk_4 {\delta(k_1+k_2-k_3-k_4)
\over (k_3-k_1)^2}
{\mbox{Tr} \over 2} \biggl[\nonumber \\
 & & \sum_{h,h'}
:\{b^\dagger_h(k_1),b_h(k_3)\}:
:\{b^\dagger_{h'}(k_2),b_{h'}(k_4)\}: \nonumber \\
&+ &{(k_1+k_3)(k_2+k_4) \over 4 \sqrt{|k_1||k_2||k_3||k_4|}}
:[a_+^\dagger(k_1),a_+(k_3)]:
:[a_{+}^\dagger(k_2),a_{+}(k_4)]: \nonumber \\
&+ &{(k_2+k_4) \over 2 \sqrt{|k_2||k_4|}} \sum_{h}
:\{ b^\dagger_h(k_1),b_h(k_3)\}:
:[a_{+}^\dagger(k_2),a_{+}(k_4)]: \nonumber \\
&+ &{(k_3+k_1) \over 2\sqrt{|k_1||k_3|}} \sum_{h'}
:[ a_+^\dagger(k_1),a_+(k_3)]::\{b^\dagger_{h'}(k_2),b_{h'}(k_4)\}:\biggr]
\label{jj}
\end{eqnarray}
The explicit form of the Hamiltonian
(\ref{jj}) in terms of the operators $b_{\pm}$,
$d_{\pm}$ and $a_{\pm}$ is straightforward to calculate, but too long
to be written down here. It should be stressed, however,
 that several $2 \rightarrow 2$ parton
processes are suppressed by a factor $1/N$, and so are ignored in
the large-$N$ limit.
No terms involving $1 \leftrightarrow 3$ parton interactions are
suppressed in this limit, however.

One  can show that this Hamiltonian  conserves total helicity $h$, which
is an additive quantum number. Moreover, the number of fermions  {\em
minus} the number of antifermions is also conserved in each interaction,
and so we have an additional quantum number ${\cal N}$.  States with
${\cal N} = even$  will be referred to as {\em boson} boundstates, while
the quantum number  ${\cal N} = odd$ will refer to {\em fermion}
boundstates. We will pay special attention to the cases ${\cal
N}= 0$ and $3$, since the associated states appear
to be  analogous to 
conventional mesons
and baryons (respectively).  

The instantaneous Coulomb interactions involving
$2 \rightarrow 2$ parton interactions behave singularly when
there is a zero exchange of momentum between identical `in'
and `out' states. The same type of
Coulomb singularity
involving $2 \rightarrow 2$ boson-boson interactions
appeared in a much simpler model \cite{dek93}, and can be shown to cancel
a `self-induced' mass term (or self-energy) obtained from normal ordering
the
Hamiltonian. The same prescription works in the model
studied here.
There are also finite residual terms left over after this cancellation
is explicitly performed for the boson-boson and boson-fermion
interactions, and they cannot be absorbed by a redefinition
of existing coupling constants.
These residual terms behave as momentum-dependent mass terms,
and in some sense represent the flux-tube energy
between adjacent partons in a color singlet state.
For the boson-boson and boson-fermion interactions they are
respectively
\begin{eqnarray}
\frac{g^2 N}{2 \pi} \cdot {\pi \over 4\sqrt{k_b k_{b'}}} \quad
\mbox{and} \quad
\frac{g^2 N}{2\pi}{1 \over k_f} \left ( \sqrt{1 +{k_f \over k_b}}-1\right)
\end{eqnarray}
where $k_b,k_b'$  denote boson momenta,
and $k_f$ denotes a fermion momentum.
These terms simply multiply the wavefunctions in the boundstate integral
equations.

\medskip

If we now include the contributions $F_h^{\dagger}
 \frac{1}{{\rm i}\partial_-} F_h$ in the light-cone Hamiltonian
(\ref{hamiltonian}), then we will
encounter another type of singularity for vanishing longitudinal
momenta $k^+=0$. This singular behavior can be shown
to cancel a (divergent) momentum-dependent mass term, which is obtained
after normal ordering the  $F_h^{\dagger}\frac{1}{{\rm i}\partial_-} F_h$
interactions and performing an appropriate (infinite)
renormalisation of the bare coupling $m_b$. This
 momentum-dependent mass term
has the explicit form
\begin{eqnarray}
\lefteqn{ \frac{s^2 g^2 N}{2 \pi} \int_0^{\infty}
dk_1 dk_2 \left\{ \frac{}{}
 \left( \frac{1}{k_2(k_1 - k_2)} + \frac{1}{k_2(k_1+k_2)} \right)
\sum_h a_{h}^{\dagger}(k_1)a_{h}(k_1) \right.}  & &
 \nonumber \\
& + & \frac{1}{k_2(k_1-k_2)} \sum_h
b_{h}^{\dagger}(k_1)b_{h}(k_1)
+  \frac{1}{k_2(k_1+k_2)}
\sum_h
d_{h}^{\dagger}(k_1)d_{h}(k_1)  \label{not2}
 \left.\frac{}{} \right\}
\end{eqnarray}
The mechanism for cancellation here is different from the Coulombic case, since
we will require specific endpoint relations relating
different wavefunctions.
Before outlining the general prescription for implementing
this cancellation, we consider
a simple rendering of the boundstate integral equations involving
the $F_h^{\dagger}\frac{1}{{\rm i}\partial_-} F_h$ interactions.
In particular, let us consider the helicity zero
sector with ${\cal N} =0$, and allow at most three partons.
Then the boundstate integral equation governing the behavior
of the wavefunction $f_{a_+ a_-}(k_1,k_2)$ for the two-boson state
$\frac{1}{N}\mbox{Tr}[a_+^{\dagger}(k_1)a_-^{\dagger}(k_2)]|0\rangle$
takes the form
\begin{eqnarray}
M^2 f_{a_+ a_-}(x_1,x_2) & = & \frac{g^2 N}{\pi} \cdot
        \frac{\pi}{4 \sqrt{x_1 x_2}}f_{a_+ a_-}(x_1,x_2)
\nonumber \\
& + & \frac{s^2 g^2 N}{\pi} \sum_{i=1,2}\int_0^{\infty}
 dy \left( \frac{1}{y(x_i - y)} + \frac{1}{y(x_i+y)} \right)
f_{a_+ a_-}(x_1,x_2) \nonumber \\
& - & msg \sqrt{\frac{N}{2\pi}} \int_0^{\infty} d\alpha d\beta \hspace{1mm}
 \delta (\alpha + \beta - x_1) \times \nonumber \\
& & \frac{1}{\sqrt{x_1}} \left( \frac{1}{\alpha} +
 \frac{1}{\beta} \right)\left[ f_{b_+ d_+ a_-}(\alpha,\beta,x_2) +
          f_{d_+ b_+ a_-}(\alpha,\beta,x_2) \right] 
\nonumber\\ & & + \dots\label{int1}
\end{eqnarray}
where $M^2 = 2P^+P^-$, and $x_i = k_i/P^+$ are  (boost invariant)
longitudinal momentum fractions.
Evidently, the integral (\ref{int1})
arising from $1 \rightarrow 2$ parton interactions
behaves singularly for vanishing longitudinal momentum
fraction $\alpha \rightarrow 0$,
or $\beta \rightarrow 0$. However, these divergences
are  precisely canceled by the momentum-dependent mass terms, which represent
the contribution (\ref{not2}).

To see this, we may consider the integral equation governing
the wavefunction $f_{b_+ d_+ a_-}(k_1,k_2,k_3)$ for the
three-parton state $\frac{1}{N^{3/2}} \mbox{Tr}[b_{+}^{\dagger}(k_1)
d_{+}^{\dagger}(k_2)a_-^{\dagger}(k_3)]|0\rangle$ :
\begin{eqnarray}
M^2 f_{b_+ d_+ a_-}(x_1,x_2,x_3) & = & m^2 \left( \frac{1}{x_1}
          + \frac{1}{x_2} \right)f_{b_+ d_+ a_-}(x_1,x_2,x_3) \nonumber \\
& + & \frac{g^2 N}{\pi} \sum_{i=1,2}
 \left[ \frac{1}{x_i} \left( \sqrt{1 + \frac{x_i}{x_3}} - 1 \right) \right]
 f_{b_+ d_+ a_-}(x_1,x_2,x_3) \nonumber \\
& - & msg \sqrt{\frac{N}{2\pi}}  \frac{1}{\sqrt{x_1+x_2}}
 \left( \frac{1}{x_1}
          + \frac{1}{x_2} \right) f_{a_+ a_-}(x_1+x_2,x_3) 
\nonumber\\ &&+ \dots
\end{eqnarray}
If we now multiply both sides of the above equation by $x_i$,
and then let $x_i \rightarrow 0$ for $i=1,2$, we deduce the relations
\begin{equation}
f_{b_+ d_+ a_-}(0,x_2,x_3) = \frac{sg}{m}\sqrt{\frac{N}{2\pi}}
\frac{f_{a_+ a_-}(x_2, x_3)}{\sqrt{x_2}}
\end{equation}
\begin{equation}
f_{b_+ d_+ a_-}(x_1,0,x_3) = \frac{sg}{m}\sqrt{\frac{N}{2\pi}}
\frac{f_{a_+ a_-}(x_1,x_3)}{\sqrt{x_1}}
\end{equation}
It is now straightforward to show that the singular behavior
of the integral (\ref{int1}) involving the wavefunction
$f_{b_+ d_+ a_-}$ may be written in terms of a
momentum-dependent
mass term involving the wavefunction $f_{a_+ a_-}$.
Similar divergent contributions are obtained from the
the wavefunctions $f_{d_+ b_+ a_-}$, $f_{a_+ b_- d_- }$ and
$f_{a_+ d_- b_- }$, all of which may be re-expressed in terms of
the wavefunction $f_{a_+ a_-}$ by virtue of corresponding
`ladder relations'.
The sum of these divergent contributions exactly
cancels the self-energy contribution.
An entirely analogous set of ladder relations
 were found for the case of fermions in the fundamental
representation of $SU(N)$ \cite{and96b}.

For the general case where states are permitted to have more than
three partons, the correct ladder relations are not immediately
obvious from an analysis of the integral equations alone.
Nevertheless, they
may be readily obtained from the constraint equation governing
the left-moving fermion field $\Psi_L$. In particular, we
have ${\rm i}\partial_- v_{\mp} = F_{\pm}$, and so
vanishing fields at spatial infinity
would imply
\begin{equation}
          \int_{-\infty}^{\infty} dx^- F_{\pm}|\Psi \rangle = 0
\end{equation}
for color singlet states $|\Psi \rangle$. The analysis
of this condition in momentum space is quite
delicate, since it involves integrals of singular
wavefunctions over spaces of measure zero \cite{abd97}.
Viewed in this way we see that
the ladder relations are the continuum equivalent of zero mode constraint
equations that
have shown to lead to spontaneous symmetry breaking in discrete light-cone
quantization \cite{bep93}.

\vskip.3in
{\bf EXACT SOLUTIONS}
\vskip.1in

For the special case
 $s=m={\tilde m}_b = 0$, the only
surviving terms in the Hamiltonian
(\ref{hamiltonian}) are the current-current interactions
$J^+ \frac{1}{\partial_-^2} J^+$ and the $\phi^4$ interaction.
This
theory  has infinitely many massless boundstates, and
the partons in these states are either
fermions or antifermions.
States with bosonic $a_{\pm}$ quanta are always
massive. One also finds that the massless states are pure,
in the sense that the number of partons is a fixed integer,
and there is no mixing between sectors of different parton number.
In particular, for each integer $n \geq 2$, one can always find a massless
boundstate consisting of a superposition
of only $n$-parton states.  A striking feature
is that the wavefunctions of these states
are {\em constant}, and so these states are natural
generalizations of the constant wavefunction solution appearing in
t'Hooft's model \cite{tho74}.

We present an explicit example below of such a constant
wavefunction solution involving a three fermion state with
total helicity $+\frac{3}{2}$, which is perhaps the simplest
case to study.
Massless states with five or more partons
appear to have more than one wavefunction which are non-zero
and constant, and in general the wavefunctions are unequal.
It would be interesting to classify all states systematically, and
we leave this to future work. One can, however, easily
count the number of massless states. In particular,
 for ${\cal N }=3$, $h=+\frac{3}{2}$ states, there is
one three-parton state, $2$ five-parton
states, $14$ seven-parton
states and $106$ nine-parton states that yield massless solutions.

Let us now consider the action of the light-cone
Hamiltonian $P^-$ on the three-parton state
\begin{eqnarray}
|b_+ b_+ b_+ \rangle &=&
\int_0^{\infty} dk_1 dk_2 dk_3
\hspace{1mm} \delta (\sum_{i=1}^3 k_i - P^+)
 f_{b_+ b_+ b_+}(k_1,k_2,k_3) \nonumber \\
&&\frac{1}{N^{3/2}} \mbox{Tr}[b_+^{\dagger}(k_1)
 b_+^{\dagger}(k_2)b_+^{\dagger}(k_3)]|0\rangle
\end{eqnarray}
The quantum number ${\cal N}$ is 3 in this case, and ensures
that the state $P^-|b_+ b_+ b_+ \rangle$ must have at least three
partons. In fact, one can deduce the following:
\begin{eqnarray}
&& P^- \mid b_+ b_+ b_+\rangle =  \int_0^{\infty} dk_1 dk_2 dk_3
\; \delta (\sum_{i\;=1}^3 k_i - P^+)  
\nonumber \\
&&
-{g^2 N \over 2\pi} \int^\infty_0 d\alpha d\beta
\frac{\delta (\alpha + \beta - k_1 -k_2)}{(\alpha - k_1)^2}  
\left[ f_{b_+ b_+b_+ }(\alpha,\beta,k_3) -  f_{b_+ b_+ b_+}(k_1,k_2,k_3) \right]
\nonumber \\
& & \frac{1}{N^{3/2}}\mbox{Tr}
\left[b_+^\dagger(\alpha)b_+^\dagger(\beta)b_+^\dagger(k_3)\right]
\mid 0 \rangle  
\nonumber 
\end{eqnarray}
\begin{eqnarray}
&&+{{g^2 N}\over 2\pi}\int^\infty_{0}
d\alpha d\beta d\gamma \sum_h
{\delta(\alpha + \beta + \gamma - k_1)
\over (\alpha+\beta)^2} f_{b_+ b_+ b_+}(\alpha + \beta + \gamma ,k_2,k_3) 
\frac{1}{N^{5/2}}\mbox{Tr}
\nonumber \\ \lefteqn{\left[
\{b_h^\dagger(\alpha),d_{-h}^\dagger(\beta)\}b_+^\dagger(\gamma)
b_+^\dagger(k_2)b_+^\dagger(k_3)-
 \{b_h^\dagger(\alpha),d_{-h}^\dagger(\beta)\}
b_+^\dagger(k_2)b_+^\dagger(k_3)b_+^\dagger(\gamma)
\right] \mid 0 \rangle}&& \nonumber \\
& & \nonumber 
\end{eqnarray}
\begin{eqnarray}
&&+{{g^2 N} \over 4\pi}\int^\infty_{0} d\alpha d\beta d\gamma
  \sum_h
{\delta(\alpha + \beta + \gamma - k_1 )
\over\sqrt{\alpha \beta}(\alpha+\beta)^2 }f_{b_+ b_+ b_+}
(\alpha + \beta + \gamma, k_2,k_3)\frac{1}{N^{5/2}}\mbox{Tr}
\nonumber \\
\lefteqn{\left[
[a_h^\dagger(\alpha),a_{-h}^\dagger(\beta)]b_+^\dagger(\gamma)
b_+^\dagger(k_2)b_+^\dagger(k_3)-
 [a_h^\dagger(\alpha),a_{-h}^\dagger(\beta)]
b_+^\dagger(k_2)b_+^\dagger(k_3)b_+^\dagger(\gamma)
\right] \mid 0 \rangle} && \nonumber \\
 &&+  \mbox{ cyclic permutations} \left. \frac{}{} \right\}
\label{exacts}
\end{eqnarray}
The five-parton states  above
correspond to virtual fermion-antifermion and boson-boson
pair creation.
The expression (\ref{exacts}) vanishes if the wavefunction
$f_{b_+ b_+ b_+}$ is constant.

\vskip.3in
{\bf CONCLUTIONS}
\vskip.1in

We have presented a non-perturbative Hamiltonian formulation of a class
of $1+1$ dimensional matrix field theories, which may be derived
from a classical dimensional reduction of
${\mbox{QCD}}_{3+1}$
coupled to Dirac adjoint fermions. We choose
to adopt the light-cone gauge $A_- = 0$, and are able
to solve numerically the boundstate
integral equations in the large-$N$ limit.
Different states may be classified according to total
helicity $h$, and the quantum number ${\cal N}$, which
defines the number of fermions minus the number of antifermions
in a state.

For a special
choice of couplings that eliminates all interactions
except those involving the longitudinal current $J^+$ and the $\phi^4$
interactions we find an
infinite number of pure massless states of arbitrary length.
The wavefunctions of these states are
always constant, and may be solved for
exactly and an example was explicitly given.
In general, a massless solution involves several
(possibly different) constant wavefunctions.
The massless solutions observed in
studies of $1+1$ dimensional
supersymmetric field theories \cite{maa95}  are not analogous to the
constant wavefunction solutions found here.

When one includes the Yukawa interactions,
singularities at vanishing longitudinal momenta
arise,
and we show in a simple case how these are canceled
by the boson and fermion self-energies.
This cancellation relies on the derivation of certain
`ladder relations', which relate different
wavefunctions at vanishing longitudinal momenta.
These relations become singular for vanishing fermion mass
$m$, and so in the context of the numerical techniques
employed here, one is prevented from studying the limit
$m \rightarrow 0$. Analytical techniques which are
currently under investigation are expected to be relevant
in this limiting case \cite{abd97}.

A particularly important property of these models is that
virtual pair creation and annihilation of bosons and fermions
is not suppressed in the large-$N$ limit, and so our
results go beyond the valence quark (or quenched) approximation.
This provides the scope for strictly field-theoretic
investigations of the internal structure of boundstates where
 `sea-quarks' and small-$x$ gluons are expected to contribute
significantly to the overall polarization of a boundstate.

The techniques employed here are not specific to the
choice of field theory, and are expected to have a
wide range of applicability, particularly in the
light-cone Hamiltonian formulation of supersymmetric
field theories.

\vskip.3in
{\bf ACKNOWLEDGMENTS}
\vskip.1in

This work was done in collaboration with Francesco
Antonuccio. The work was supported in part by a grant from the US
Department of Energy. Travel support was provided in part by a
NATO collaborative grant.

\end{document}